\documentclass[aps,preprint,nofootinbib,preprintnumbers]{revtex4}

\usepackage{amsmath,amstext,amssymb}
\usepackage{comment,url}
 
\newcommand{\beq}{\begin{equation}}
\newcommand{\eeq}{\end{equation}}

\begin{document}

\preprint{UW-NT-13-27}

\author{Matthias Kaminski}
\email{mski@uw.edu}
\author{Sergej Moroz}
\email{morozs@uw.edu}
\affiliation{Department of Physics, University of Washington, Seattle, WA 98195, USA }

\title{Non-Relativistic Parity-Violating Hydrodynamics\\ 
in Two Spatial Dimensions\footnote{Authors in alphabetical order.}}
\begin{abstract}
We construct the non-relativistic parity-violating hydrodynamic description of a two-dimensional dissipative, normal fluid in presence of small $U(1)$ background fields and vorticity. This is achieved by taking the non-relativistic limit of the recently developed relativistic hydrodynamics in $2+1$ dimensions. We identify and interpret the resulting parity-violating contributions to the non-relativistic constitutive relations, which include the Hall current flowing perpendicular to the temperature gradient, the Hall viscosity and the Leduc-Righi energy current.  Also a comparison of our findings is made with the non-relativistic parity-violating hydrodynamics obtained from a light-cone dimensional reduction.
\end{abstract}

\date{\today}

\maketitle


\section{Introduction} \label{sec:introduction}

Hydrodynamics is a universal language we use to describe dynamics of interacting many-body systems in the limit of low energies and long wave lengths. Although hydrodynamics is an old subfield of physics, a lot of progress in its development has been achieved during recent years. To a large extend these advances were triggered by new experiments investigating both extremely high-energy (relativistic heavy ion collisions) and low-energy (ultracold quantum liquids) states of matter.

Many-body systems living in two spatial dimensions are abundant in condensed matter physics. Among these parity-violating fluids play a prominent role.\footnote{In two spatial dimensions parity is a discrete transformation that inverts one of the spatial coordinates.} 
Here we will consider parity-violating effects in the non-relativistic fluid composed of single species of massive particles.
Such a fluid has a $U(1)$ internal symmetry associated with particle number conservation.\footnote{This internal $U(1)$ symmetry may be global or local. Our considerations in this paper apply to both of these cases. We will refer to the sources coupling to the current associated with this $U(1)$ symmetry as electric and magnetic fields. Note that we set the $U(1)$ charge of a single particle to one.}
The goal of this paper is to develop non-relativistic planar hydrodynamics of such a parity-violating fluid in the normal, i.e. non-superfluid regime, at vanishing background magnetic field and vorticity. However, we do take into account effects due to vorticity and magnetic field which are of first order in derivatives of hydrodynamic variables, as we explain below. 

Here we list physical systems, where this version of hydrodynamics might be useful:
\begin{itemize}
 \item{\bf Fluids of chiral molecules:}   Chiral molecules are not identical to their mirror image and thus microscopically break parity. While chiral hydrodynamics in three spatial dimensions was studied before(see \cite{Andreev:2010} and references therein), our theory should be applicable to two-dimensional thin films of fluids composed of chiral molecules.
 \item{\bf Fluids of spinning particles:} Spin is a pseudoscalar in two spatial dimensions and thus an ensemble of spin-polarized particles breaks parity.  
 \item {\bf Two-dimensional systems with Rashba spin-orbit coupling:} In the presence of the external electric field $\mathbf{E}=\mathcal{E} \mathbf{z}$ the Rashba spin-orbit coupling $H_{R}^{(3d)}\sim \mathbf{S}\cdot (\mathbf{p}\times \mathbf{E})$ reduces to the two-dimensional form $H_{R}^{(2d)}\sim \mathcal{E} (S_x p_y-S_y p_x)$.   Since this coupling acts as an effective momentum-dependent magnetic field, it breaks parity in two spatial dimensions.    
 Two-component Fermi gas with the Rashba spin-orbit coupling of the form $H^{(2d)}_{R}\sim \sum_{\mathbf{k}} \psi(\mathbf{k})^{\dagger}(\sigma_x p_y-\sigma_y p_x)\psi(\mathbf{k})$ can be now realized in experiments with ultracold atoms \cite{2013arXiv1308.6533G,DellAnna2011}. Our hydrodynamics should be applicable in the normal phase of this system.
\item {\bf Non-relativistic anyons:} Consider non-relativistic massive particles minimally coupled to a statistical Chern-Simons gauge field. Generically such a system exhibits anyonic quasiparticles. Due to the Chern-Simons term, parity is broken explicitly in this non-relativistic model of anyons. At low temperatures the anyon fluid is supposed to be superfluid and can be described with hydrodynamics of \cite{PhysRevB.41.240}. We expect our equations to be applicable in the normal (high-temperature) phase of the anyon fluid. 
\end{itemize}

Our main motivation for the derivation presented in this paper is the assumption that every system on a microscopic level has to obey relativistic symmetries. Furthermore, we wish to find a low-energy effective description of systems like those mentioned above. Therefore, our starting point is relativistic hydrodynamics. Taking the speed of light to infinity we obtain a non-relativistic version of hydrodynamics consistent with non-relativistic general coordinate invariance, as discussed below and in Section~\ref{sec:derivationsNRPVH}. One of the virtues of this approach is its systematic character. 
When deriving the relativistic equations one can exploit Lorentz invariance in order to obtain all the terms allowed by symmetry: one constructs all the Lorentz covariant quantities which can be formed from index contractions among the hydrodynamic variables and their derivatives~\cite{LL6}, see~\cite{Jensen:2011xb} for the parity-violating case in $2+1$ dimensions. This systematic approach in particular ensures that no terms are missed and indeed the most general, complete version of relativistic hydrodynamics is found. An analogous systematic construction principle utilizing non-relativistic symmetries is currently not known, although we propose an approach in Section~\ref{sec:derivationsNRPVH}. Another virtue of our approach is the possibility that it may be more restrictive. While traditional derivations of non-relativistic hydrodynamics impose Galilean invariance, our derivation imposes a potentially more restrictive invariance under non-relativistic general coordinate transformations. As a result of our analysis we find relations for various transport coefficients which to our knowledge are new. For example, our derivation implies that a particular contribution to the energy current perpendicular to a temperature gradient can be expressed in terms of the equilibrium pressure, non-relativistic energy density, and the particle density. Similarly, we are able to determine the shift in the pressure due to fluctuations in the magnetic field and in the vorticity.
Finally, there are some issues which are easily addressed in the relativistic theory, but which are obscured in the non-relativistic context. One example for this is the issue of hydrodynamic frames and the transformations between them, which we discuss in Section~\ref{sec:hydroFrames}. 

The paper is structured as follows: In the next section, Section \ref{sec:derivationsNRPVH},  we consider various derivations of non-relativistic hydrodynamics.
 In Section \ref{sec:RPVH} we  start from the relativistic parity-violating hydrodynamics in $2+1$ dimensions that was constructed recently to first order in derivatives in \cite{Jensen:2011xb}. Subsequently, in Section \ref{sec:NRPVH} we perform the non-relativistic limit, i.e.,  send $c\to \infty$, and obtain the non-relativistic version of parity-violating hydrodynamics. In the parity-preserving case we recover the non-relativistic dissipative (Navier-Stokes) hydrodynamics written in textbooks.  On the other hand, if parity is broken the non-relativistic constitutive relations are modified. We identify these modifications and uncover their physical meaning. Frame transformations are discussed in Section \ref{sec:hydroFrames} and we draw conclusions in Section \ref{sec:conclusions}.

\section{Derivations of Non-Relativistic Hydrodynamics} \label{sec:derivationsNRPVH}

There are various distinct ways to derive non-relativistic hydrodynamics. In this section we briefly describe three of these possibilities in order to contrast our approach from others.

\subsection{$c\to\infty$ limit of Relativistic Hydrodynamics}
In this work we adopt the point of view that any truly microscopic theory has to be relativistic. This means that these microscopic theories have to obey Lorentz symmetry. 
Any effective theory, such as hydrodynamics, describing the same system in some regime has to obey Lorentz symmetry as well. Now we choose to observe the system at velocities which are much smaller than the speed of light. Therefore, the correct non-relativistic version of hydrodynamics should emerge from relativistic hydrodynamics in the limit of infinite speed of light $c\to \infty$. This is exactly the procedure we will carry out in this paper.

Before getting started, let us mention that there is a caveat to the above arguments. We have tacitly assumed that the $c\to \infty$ limit of the relativistic effective theory is identical to the effective theory of the $c\to \infty$ limit of the microscopic theory. In other words we have assumed that the two limits commute, i.e.,
\begin{eqnarray}
\boxed{\text{relativistic microscopic theory}} & 
\overset{\tiny \text{hydrod. limit}}{
\longrightarrow
} 
& \boxed{\text{relativistic hydrodynamics}} \nonumber \\
c\to \infty 
\downarrow\hphantom{hallohallo} & & \hphantom{hallohallo} \downarrow 
c\to \infty  
\nonumber \\
\boxed{\text{non-relativistic microscopic theory}} & 
\overset{\tiny \text{hydrod. limit}}{
\longrightarrow
} 
& \boxed{\text{non-relativistic hydrodynamics}} \nonumber \, 
\end{eqnarray}
Although we do not prove the commutativity here, in the following subsection we propose an approach which may answer this question. 

\subsection{General Coordinate Invariance}
Probably the most rigorous derivation of the most general non-relativistic hydrodynamics should make use of general coordinate invariance~\cite{Son:2005rv}. Here we would not have to rely on limits or their commutativity. 
As we will review in section~\ref{sec:RPVH}, the equivalent approach exploiting the relativistic general coordinate invariance has been extremely useful in deriving the most-general version of relativistic hydrodynamics in 2+1 dimensions~\cite{Jensen:2011xb}.  

Non-relativistic general coordinate invariance is a symmetry with respect to non-relativistic diffeomorphisms and abelian gauge transformations associated with particle number phase rotation.
It was shown in~\cite{Son:2005rv} that it can be obtained as the non-relativistic limit of a relativistic general coordinate invariance. Non-relativistic general coordinate invariance can be looked upon as a local version of the Galilean invariance, and thus it is more restrictive than Galilean symmetry.

In order to find the most general constitutive equations of hydrodynamics, one has to write down all possible contributions to energy-momentum tensor and conserved currents which are covariant under non-relativistic general coordinate transformations. These expressions are then restricted afterwards by use of the positivity of the local entropy production as explained in standard textbooks such as~\cite{LL6}.  For the parity-invariant case this approach was carried out in \cite{Son:2007}.

\subsection{Light-cone Dimensional Reduction of Relativistic Hydrodynamics}

Light-cone dimensional reduction (LCDR) is an alternative path to non-relativistic hydrodynamics in two spatial dimensions. In this framework one starts with the relativistic fluid in 3+1 dimensional Minkowski spacetime parametrized by the light-cone coordinates $\{x^+, x^-, x^i \}$. Then, by imposing that all hydrodynamic variables are independent of the light-cone coordinate $x^-$, the non-relativistic hydrodynamic equations in two spatial dimensions are obtained. The light-cone coordinate $x^+$ plays the role of the time in the non-relativistic theory. LCDR of the ideal and viscous parity-invariant conformal hydrodynamics was performed in \cite{Rangamani:2008gi}. The extension to the charged parity-violating conformal case was undertaken in \cite{Brattan:2010bw}. 

It is worth emphasizing that the hydrodynamic transport coefficients of the non-relativistic fluid obtained from LCDR are not the most general ones allowed by symmetries, but obey additional constraints. For instance, it was found in \cite{Rangamani:2008gi} that the Prandtl number 
of the non-relativistic conformal fluid obtained from LCDR is always equal to one. In Appendix \ref{sec:dimensionalReduction} we identify some of the LCDR constraints on the transport coefficients in the parity violating sector. As a result, LCDR is a viable method to construct some version of (parity-violating) non-relativistic hydrodynamics in two spatial dimensions. It remains to be seen, however, whether the LCDR constraints on the transport coefficients have some deeper physical meaning or if they are just artifacts of this particular method.

\section{Relativistic Parity-Violating Hydrodynamics} \label{sec:RPVH}
From a modern point of view hydrodynamics is an expansion in gradients of hydrodynamic variables, such as temperature, fluid velocity, and chemical potential; or gradients of external sources. Gradients in Fourier space can be rewritten in terms of frequency $\omega$ and momentum $k$, such that in relativistic hydrodynamics $\omega, k\ll T$. \footnote{Note that this expansion can be always carried out. The question of when this is a reliable effective description of the underlying system is separate. For a recent discussion of the applicability of hydrodynamics see for example~\cite{Schafer:2009dj}.}

In this section we review the most general formulation of the theory of relativistic hydrodynamics with a $U(1)$ conserved current. By this we mean the form of the so called constitutive equations
\begin{eqnarray}\label{eq:generalConstitutive}
T^{\mu\nu} &=&  T^{\mu\nu} (T, \mu, u^\mu; A^\mu) ,\\
J^\mu & = & J^\mu (T, \mu, u^\mu; A^\mu) \, ,
\end{eqnarray}
and the corresponding hydrodynamic conservation equations encoding the flow of energy, momentum, and charge
\begin{eqnarray}
\nabla_\mu T^{\mu\nu} &=& F^{\nu\mu} J_\mu \, ,\\
\nabla_\mu J^\mu & = & 0 \, .
\end{eqnarray}
Here Greek indices run over the 2+1 spacetime dimensions, $\mu, \nu, \dots = 0,1,2$ and $\nabla_{\mu}$ stands for the covariant derivative.
The constitutive equations~\eqref{eq:generalConstitutive} express the relativistic energy-momentum tensor $T^{\mu\nu}$ and the $U(1)$ conserved current $J^\mu$ in terms of the hydrodynamic variables: temperature $T(x)$, chemical potential $\mu(x)$, and fluid velocity $u^\mu(x)$; and the external electromagnetic source $A^\mu(x)$. These are all functions of the spacetime coordinates $x=(t,\,x_1,\,x_2)$. These equations are invariant under field redefinitions of the hydrodynamic variables $T\to T+\delta T,\, \mu \to \mu+\delta \mu,\, u^\mu\to u^\mu+\delta u^\mu$. This field ambiguity means that the constitutive equations can be written in different forms which are knows as hydrodynamic frames. We will discuss this fact in more detail in section~\ref{sec:hydroFrames}. 

\paragraph{Constitutive Equations with $\Omega$ and $B$.}
Our goal in this section is to describe a charged relativistic normal fluid in 2+1 dimensions. The most general constitutive equations of parity-violating hydrodynamics in $2+1$ dimensions with one conserved current $J^\mu$ were obtained to first order in derivatives in~\cite{Jensen:2011xb}. 
This derivation allows for a small\footnote{Small denotes that $B$ and $\Omega$ are built from first derivatives acting on hydrodynamic variables or on the external source $A^\mu$. 
} magnetic field $B$ and small vorticity $\Omega$. Both $\Omega$ and $B$ are pseudoscalars in 2+1 dimensions. In the so-called magnetovortical frame\footnote{\label{footnote} It is important to note that in the parity-invariant case, i.e.,  $\tilde \chi_T=\tilde \chi_E= 
\tilde x_\Omega = \tilde x_B=\tilde\eta=\tilde \sigma=0$ the magnetovortical frame is equivalent to the well-known Landau frame.} one finds\begin{subequations}
\label{E:T1J1L}
\begin{eqnarray}
  &&   T^{\mu\nu} = \left( \epsilon - \tilde{x}_\Omega \Omega \right )  u^\mu u^\nu
  			    +\left(P - \zeta \nabla_\alpha u^\alpha - \tilde{x}_B B - \tilde{x}_\Omega \Omega\right) \Delta^{\mu\nu}
			    -\eta \sigma^{\mu\nu} - \tilde{\eta} \tilde{\sigma}^{\mu\nu} \,,  \\
  &&  J^\mu = \left (n -\tilde{x}_B \Omega \right )  u^\mu + \sigma V^{\mu} + \tilde{\sigma} \tilde{V}^{\mu} + \tilde{\chi}_E \tilde{E}^{\mu} + \tilde{\chi}_T \epsilon^{\mu\nu\rho}u_{\nu} \nabla_{\rho} T \, ,
\end{eqnarray}
\end{subequations}
where $\epsilon$ is the relativistic energy density, $P$ is the pressure, $n$ is the relativistic charge density, and the well-known transport coefficients of electrical conductivity, shear viscosity and bulk viscosity are given by $\sigma$, $\eta$, and $\zeta$, respectively. In the following we distinguish hydrodynamic transport coefficients from thermodynamic transport coefficients, see~\cite{Jensen:2011xb} for a detailed discussion. The thermodynamic transport coefficients $ \tilde{\chi}_T,\, \tilde{\chi}_E $ and $\tilde{x}_\Omega,\, \tilde{x}_B$ are associated with parity-violation and are known exactly in terms of derivatives of the thermodynamic quantities $P, \epsilon, n$.\footnote{Compared to \cite{Jensen:2011xb}, here  the charge density in the rest frame will be denoted by $n$. Otherwise, in this section we follow the conventions of \cite{Jensen:2011xb} and have $\eta_{\mu\nu}=(-++)$, $\epsilon^{012}=\epsilon^{12}=+1$, $c=1$.} In addition, $\tilde \eta$ and $\tilde\sigma$ are two parity-violating hydrodynamic transport coefficients that are not fixed by thermodynamics.
The other quantities appearing in the constitutive relations (\ref{E:T1J1L}) are
\begin{subequations}
\label{E:defs}
\begin{align}
\label{E:OandB}
	 & \Omega = -\epsilon^{\mu\nu\rho}u_{\mu} \nabla_{\nu} u_{\rho}, &
	& B = -\frac{1}{2} \epsilon^{\mu\nu\rho}u_{\mu} F_{\nu\rho}, \\
	& E^{\mu}  =  F^{\mu\nu}u_{\nu}, &
	& V^{\mu}  = E^{\mu} - T \Delta^{\mu\nu}\nabla_{\nu} \frac{\mu}{T}, \\
\label{E:sigmaDef}
	& \Delta^{\mu\nu} = u^{\mu}u^{\nu} + \eta^{\mu\nu}, &
	& \sigma^{\mu\nu}  = \Delta^{\mu\alpha} \Delta^{\nu\beta} \left(\nabla_{\alpha}u_{\beta} + \nabla_{\beta} u_{\alpha} - g_{\alpha\beta} \nabla_{\lambda} u^{\lambda} \right) ,
\intertext{and}
	&\tilde{E}^{\mu}  = \epsilon^{\mu\nu\rho}u_{\nu}E_{\rho}\,,&
	&\tilde{V}^{\mu}  = \epsilon^{\mu\nu\rho}u_{\nu} V_{\rho}\,, \\
	&\tilde{\sigma}^{\mu\nu} = \frac{1}{2} \left( \epsilon^{\mu\alpha\rho} u_{\alpha} \sigma_{\rho}^{\phantom{\rho}\nu} +  \epsilon^{\nu\alpha\rho} u_{\alpha} \sigma_{\rho}^{\phantom{\rho}\mu} \right)\,.&
\end{align}
\end{subequations}

These constitutive equations were found in~\cite{Jensen:2011xb} following a two step procedure. The first step is to write down all possible vectors and tensors which can be constructed from the hydrodynamic variables and the external source using single derivative and the completely antisymmetric tensor $\epsilon^{\mu\nu\rho}$. In the second step these vectors and tensors are restricted to the ones appearing in the constitutive relations~\eqref{E:T1J1L}. The two alternative ways to achieve this restriction are: (i) follow Landau's entropy argument and require the local entropy production to be positive~\cite{LL6}; (ii) analyze the hydrodynamic two-point functions and derive restrictions from basic field theoretic principles such as unitarity (see~\cite{Jensen:2011xb} and~\cite{Jensen:2012jh,Banerjee:2012iz} for details).

\paragraph{Parity-odd Thermodynamic and Hydrodynamic Transport Coefficients.}
The constitutive equations~\eqref{E:T1J1L} contain the ideal fluid parts $T^{\mu\nu}_\text{ideal} = \epsilon u^\mu u^\nu +P \Delta^{\mu\nu}$ and $J^{\mu}_\text{ideal} = n u^\mu$. But more interestingly the equations~\eqref{E:T1J1L} describe the response of the current $J^\mu$ and energy momentum tensor $T^{\mu\nu}$ to gradients of the hydrodynamic variables and external sources. 
The strength of this response is encoded in transport coefficients. The parity-even transport properties encoded in Eq. \eqref{E:T1J1L} are well-know from textbooks \cite{LL6}.
Let us now briefly describe the most interesting (previously neglected) parity-odd contributions. In our notation these transport coefficients  are marked with a tilde. 
The parity-odd cousin of the electric field $\tilde E^\mu$ appears in Eq. \eqref{E:T1J1L} with two distinct coefficients $\tilde \chi_E$ and $\tilde \sigma$. Together they contribute to the anomalous Hall conductivity~\cite{Jensen:2011xb}. Correspondingly, the parity-odd cousin of the shear viscosity $\tilde\eta$ is the Hall viscosity introduced in \cite{Avron:1995fg,1997physics..12050A} (see also recent works \cite{2009PhRvB..79d5308R,2011PhRvB..84h5316R,Hoyos:2011ez}). Finally, the Hall response of the current $J^{\mu}$ to a temperature gradient appears in Eq. \eqref{E:T1J1L} \cite{Ryu:2011wq}. This is proportional to a separate transport coefficient $\tilde \chi_T$.

\paragraph{Thermodynamic Equilibria with Nonzero $\Omega$ and $B$.}
 As argued in~\cite{Jensen:2011xb, Jensen:2012jh} neither magnetic field nor vorticity contribute to entropy production. Furthermore, a (small) constant magnetic field and vorticity do not push the system out of equilibrium. 
Hence both vorticity and magnetic field can be used to label equilibrium states. 
Consequently all thermodynamic quantities such as pressure $P$, energy density $\epsilon$ or entropy density $s$ are functions not only of temperature $T$ and chemical potential $\mu$, but also of the vorticity $\Omega$ and magnetic field strength $B$. For the pressure $P(T,\, \mu,\, B,\, \Omega)$ evaluated at $B=\Omega=0$ this implies in particular for the total differential
\begin{equation}\label{eq:GibbsDuhemRelation}
dP = s dT + n d\mu + \frac{\partial P}{\partial B} dB + \frac{\partial P}{\partial \Omega} d\Omega\, ,
\end{equation}
where the partial derivatives are evaluated at $B=0$ and $\Omega=0$. Note, however, that the relation $\epsilon+P=sT+n \mu$ remains unmodified since we study only equilibria with $B=\Omega=0$ in this paper. 
The non-zero $B$ and $\Omega$ appearing in the constitutive equations \eqref{E:T1J1L} can be regarded as small perturbations around these equilibrium states. These perturbations in the magnetic field and in the vorticity are of the same order in the gradient expansion as the first order gradients of temperature, chemical potential, and fluid velocity appearing in the constitutive relations \eqref{E:T1J1L}.

The perturbations in $B$ and $\Omega$ in the constitutive relations break parity extrinsically. Note however, that some of the parity-violating transport effects are present even when the perturbations in magnetic field $B$ and vorticity $\Omega$ vanish explicitly in the constitutive equations \eqref{E:T1J1L}. For instance the current $J^\mu$ responds to a temperature gradient with the contribution $\tilde\chi_T \epsilon^{\mu\nu\rho} u_\nu \nabla_\rho T$ in equation \eqref{E:T1J1L}. This shows that the fluid under consideration has to break parity intrinsically. On the level of our effective hydrodynamic description this intrinsic parity-violation is encoded in the transport coefficients $\tilde\eta$, $\tilde\sigma$ and thermodynamic transport coefficient $\tilde \chi_T$.

\section{Non-Relativistic Limit} \label{sec:NRPVH}
In this section we take the non-relativistic limit of the constitutive and hydrodynamic equations reviewed in the previous section. For this purpose we first consider the well-known parity symmetric case, and then discuss separately the parity-violating contributions. In this section we will restrict our analysis to flat spacetime.

\subsection{Parity-Preserving Case}
Taking the non-relativistic limit involves splitting the relativistic energy density in the rest frame into the rest-mass and the internal part
\begin{equation}
\epsilon \to n m c^2+ \epsilon_{nr}\, ,
\end{equation}
where $n$ denotes the number density of particles of mass $m$ in the rest frame. In similar spirit, the relativistic chemical potential $\mu$ decomposes as
\beq
\mu\to mc^2+\mu_{nr},
\eeq
which defines the non-relativistic chemical potential $\mu_{nr}$.
Furthermore, we have to reinstate powers of $c$ in
\begin{equation}
\begin{split}
u^{\mu} &\to \Gamma (1, v^i/c)\, , \\ 
\partial_\mu &\to (\partial_t/c, \partial_i)\, , \\
V^{\mu}&\to \frac{1}{c}E^{\mu} -\frac{1}{c^2} T \Delta^{\mu\nu}\partial_{\nu} \frac{\mu}{T}\, , \\
F^{\nu\mu}J_\mu&\to c F^{\nu\mu}J_\mu, 
\end{split}
\end{equation}
where $\Gamma=\frac{1}{\sqrt{1-v^2/c^2}}$. In addition, it is convenient to express our findings in terms of the rest-mass energy density in the moving frame which is given by $\rho c^2=\Gamma n m c^2$ (see \cite{Koide2013} for an alternative viewpoint).\footnote{In this paper we perform the non-relativistic limit of the charged fluid of massive particles. This should be contrasted to the non-relativistic scaling limit of the conformal fluid of massless particles (e.g. photon gas) performed recently in \cite{Fouxon2008,Bhattacharyya:2008kq,Brattan:2011my,Bagchi:2009my}.}

The non-relativistic limit of the parity-invariant part of the energy-momentum tensor $T^{\mu\nu}$ in Eq. \eqref{E:T1J1L} reads
\begin{equation} \label{NRlim}
 \begin{split}
T^{00} &= c^2\rho+ \left[\epsilon_{nr} + \frac{1}{2} \rho v^2 \right]+\mathcal{O}(1/c^2)\, , \\
T^{0i} &= c\rho v^i+ \frac{1}{c} \underbrace{\left[ (w_{nr}+\frac{1}{2} \rho v^2)v^i -\Sigma^{ij} v_j \right]}_{j_{\epsilon}^i-j_{\epsilon,th}^i} +\mathcal{O}(1/c^3)\, , \\
T^{ij} &= \underbrace{P\delta^{ij}+ \rho v^i v^j-\Sigma^{ij}}_{\Pi^{ij}} +\mathcal{O}(1/c^2)\, ,
\end{split}
\end{equation}
with
\beq
w_{nr}=\epsilon_{nr}+P,
\eeq
\beq
\Sigma^{ij}=\eta_{nr}(\partial^i v^j+\partial^j v^i-\delta^{ij}\partial_k v^k)+\zeta_{nr}\delta^{ij}\partial_k v^k,
\eeq
where we introduced the non-relativistic shear viscosity $\eta_{nr}=\eta/c$ and bulk viscosity $\zeta_{nr}=\zeta/c$ assuming they are finite in the non-relativistic limit \cite{LL6}.

In a similar fashion we find the non-relativistic limit of the parity-invariant part of the $U(1)$ symmetry current $J^{\mu}$ in Eq. \eqref{E:T1J1L}
\begin{equation}
\begin{split} \label{current}
J^0&=\frac{\rho}{m}+\mathcal{O}(1/c^4) \, , \\
J^i&=\frac{1}{c}\frac{\rho v^i}{m} + \frac{1}{c^3}\frac{1}{m} \underbrace{\kappa_{nr}\partial^i T}_{-j^i_{\epsilon, th}} +\mathcal{O}(1/c^5)\, .
\end{split}
\end{equation}
Following \cite{LL6} we introduced the non-relativistic thermal conductivity $\kappa_{nr}=\lim_{c\to\infty}\frac{\sigma}{c} \frac{(\epsilon+P)^2}{n^2 T}$ and assumed that it is finite.\footnote{We find $\kappa_{nr}/\sigma_{nr}=m^2/T$, where $\sigma_{nr}=\sigma c^3$ was introduced. Due to Galilean invariance the heat conductivity $\kappa_{nr}$ is proportional the charge conductivity $\sigma_{nr}$. In hydrodynamics, the proportionality factor however should be contrasted with the celebrated Wiedemann-Franz law which predicts $\kappa_{nr}/\sigma_{nr}\sim T$.} 

Now we are ready to demonstrate how the relativistic hydrodynamic equations
\beq \label{relcons}
\nabla_{\mu}T^{\mu\nu}=F^{\nu\mu}J_{\mu}\, , \quad \nabla_{\mu} J^{\mu}=0 \, ,
\eeq
reduce in the non-relativistic limit to the well-known equations of the non-relativistic dissipative (Navier-Stokes) hydrodynamics. One finds the particle mass density, momentum and energy hydrodynamic equations (see e.g. \cite{LL6})
\beq \label{conserv}
\boxed{
\begin{split}
&\partial_t \rho+\partial_i (\rho v^i)=0, \\
&\partial_t (\rho v^i)+\partial_j \Pi^{ij}=\frac{\rho}{m}\left(\mathcal{E}^i+\mathcal{B}\epsilon^{ij}v_j \right), \\
&\partial_t \left( \epsilon_{nr} + \frac{1}{2} \rho v^2 \right)+\partial_i j_{\epsilon}^i=\frac{\rho}{m}\mathcal{E}^i v_i, \\
\end{split}
}
\eeq
where $\mathcal{B}=F^{12}$ and $\mathcal{E}^i=cF^{0i}$ denote the magnetic and electric fields respectively.
The first equation requires the conservation of mass (continuity equation). The second equation governs the time evolution of momentum in the presence of external electromagnetic fields (Lorentz force). These equations are obtained from taking the limit $c\to\infty$ of the equation $\nabla_\mu T^{\mu\nu}=F^{\nu\mu}J_{\mu}$. Obviously, in this limit $\nabla_\mu J^\mu=0$ becomes redundant since we are dealing with a one-component fluid with a fixed charge-to-mass ratio. The third equation that dictates the time evolution of the non-relativistic energy (including Joule heating) is obtained by subtracting the rest mass energy current from the total energy current, i.e.,
\beq \label{Son}
\nabla_\mu (T^{\mu0}-mc^2 J^\mu)=F^{0\mu}J_\mu. 
\eeq

In summary, the relativistic viscous hydrodynamics formulated in the Landau frame reduces to its non-relativistic textbook version in the limit $c\to\infty$.  We will demonstrate in section \ref{sec:hydroFrames} that by taking the non-relativistic limit of the viscous hydrodynamics formulated in the Eckart frame one also recovers the canonical constitutive relations of the Navier-Stokes non-relativistic hydrodynamics \cite{Kovtun:2012rj}.


In the absence of the external electromagnetic sources the conservation equations \eqref{relcons} have to be satisfied order by order in the $1/c^2$ expansion. In other words, they give rise to (generically) an infinite number of conservation equations for the expansion coefficients of $J^\mu$ and $T^{\mu\nu}$ of the non-relativistic expansion. By examining Eqs. \eqref{NRlim} and \eqref{current}, it is obvious that the leading-order coefficients satisfy the conservation laws. At the next-to-leading order, however, the situation becomes subtle in the Landau frame and it requires clarification. 
Neither our derivation nor any of our results depend on these subtleties. Therefore we leave this issue for future investigation.

\subsection{Parity-Violating Case}
Following the same steps as in the previous subsection, the non-relativistic limit of the parity-violating part of the constitutive relations can be found. First we reinstate powers of $c$ in the parity-violating sector
\beq
n-\tilde x_B \Omega\to n-\frac{1}{c}\tilde x_B \Omega\, .
\eeq
Guided by the parity-invariant dissipative hydrodynamics, we will introduce the non-relativistic parity-violating transport coefficients $\tilde{\eta}_{nr}$ and $\tilde{\kappa}_{nr}$. On the other hand,  for the thermodynamic transport coefficients $\tilde{x}_{B}$, $\tilde{x}_{\Omega}$, $\tilde{\chi}_{E}$ and $\tilde{\chi}_{T}$ derived in  \cite{Jensen:2011xb}\begin{eqnarray}
\tilde{x}_{B}&=& \frac{\partial P}{\partial B}, \\
\tilde{x}_{\Omega}&=& \frac{\partial P}{\partial \Omega}, \\
\tilde{\chi}_{E}&=& \frac{\partial n }{\partial B}+\frac{n_0}{\epsilon_0+P_0}\left[\frac{\partial P}{\partial B}-c\frac{\partial n}{\partial \Omega} \right], \\
T \tilde{\chi}_{T}&=&  \frac{1}{c}\frac{\partial \epsilon }{\partial B}+\frac{n_0}{\epsilon_0+P_0}\left[\frac{\partial P}{\partial \Omega}-\frac{\partial \epsilon}{\partial \Omega} \right] \, ,
\end{eqnarray}
the leading terms in the non-relativistic expansion read\footnote{Subscript $_0$ emphasizes that a given thermodynamic function is evaluated at $\mathcal{B}=\Omega_{nr}=0$. Thus by definition for any function $f_0$ one finds $\frac{\partial f_0}{\partial \mathcal{B}}=\frac{\partial f_0}{\partial \Omega_{nr}}=0$.}
\begin{eqnarray}
\tilde{x}_{B}&\rightarrow& \frac{\partial P}{\partial \mathcal{B}}, \\
\tilde{x}_{\Omega}&\rightarrow& c\frac{\partial P}{\partial \Omega_{nr}},  \\
\tilde{\chi}_{E}&\rightarrow&\frac{1}{m} \left[\frac{\partial }{\partial \mathcal{B}}-\frac{1}{m}\frac{\partial }{\partial \Omega_{nr}} \right](\rho-\frac{1}{2} \frac{\rho v^2}{c^2})+\frac{1}{mc^2}\left[\frac{\partial P}{\partial \mathcal{B}} +\frac{w_{nr,0}}{m \rho_0}\frac{\partial \rho}{\partial \Omega_{nr}} \right]= \frac{1}{mc^2}\frac{\partial \Pi}{\partial \mathcal{B}} , \\
T \tilde{\chi}_{T}&\rightarrow& \frac{1}{c} \left[\frac{\partial }{\partial \mathcal{B}}-\frac{1}{m}\frac{\partial }{\partial \Omega_{nr}} \right](\rho c^2+\epsilon_{nr}-\frac{1}{2}\rho v^2)+\frac{1}{mc}\left[ \frac{\partial P}{\partial\Omega_{nr}}+\frac{w_{nr,0}}{\rho_0}\frac{\partial \rho}{\partial \Omega_{nr}}\right] \\
&=&\frac{1}{mc}\frac{\partial \Pi}{\partial\Omega_{nr}}, \nonumber
\end{eqnarray}
where the non-relativistic vorticity is given by $\Omega_{nr}=\epsilon^{ij}\partial_{i} v_j$ and we introduced $\Pi=P+w_{nr,0}\ln \rho$.
In addition, we used the equivalence of the Lorentz and Coriolis forces in the underlying microscopic non-relativistic theory. In the underlying microscopic non-relativistic theory of our fluid the particles simultaneously carry both, mass and a $U(1)$ charge. Hence, if exposed to a magnetic field and vorticity simultaneously these particles will experience both, the Coriolis force and the Lorentz force. As a consequence, we assume that the microscopic theory depends on $B$ and $\Omega$ only in the combination $\mathcal{B}+m \Omega_{nr}$. One may think of this property as a symmetry of the microscopic theory which ties the magnetic field to the vorticity. Alternatively, this observation can be interpreted as a consequence of the non-relativistic general coordinate transformation introduced in \cite{Son:2005rv}. Indeed, while neither the magnetic field $\mathcal{B}$ nor the non-relativistic vorticity $\Omega_{nr}$ are scalars under these transformations, the linear combination $\mathcal{B}+m \Omega_{nr}$ transforms as a scalar.  
Any effective description, in particular our non-relativstic hydrodynamics, has to obey the same symmetries as the underlying microscopic theory.
Therefore we assumed that thermodynamic and hydrodynamic quantities depend only on the linear combination $\mathcal{B}+m \Omega_{nr}$.\footnote{This might be only true to linear order in $\mathcal{B}$ and $\Omega_{nr}$. Indeed, for a non-relativistic system, the Hamiltonian $H_{\omega}$ written in the frame rotating with the angular velocity $\omega=\Omega_{nr}/2$ and the Hamiltonian $H_{\mathcal{B}}$ in the presence of the constant magnetic field $\mathcal{B}=m\Omega_{nr}$ differ from each other by a term $\sim \Omega_{nr}^2 \mathbf{x}^2$. This quadratic term, however, does not modify the non-relativistic expansion of the parity-odd thermodynamic transport coefficients.} Hence in the following $\frac{\partial}{\partial \mathcal{B}}=\frac{1}{m}\frac{\partial}{\partial \Omega_{nr}}$.

For the parity-violating part of the energy-momentum tensor $T^{\mu\nu}$ in Eq. \eqref{E:T1J1L} we obtain
\begin{equation} \label{NRlimpv}
 \begin{split}
T^{00}_{\not P} &=-\frac{\partial P}{\partial \Omega_{nr}}\Omega_{nr}+\mathcal{O}(1/c^2)\, , \\
T^{0i}_{\not P} &= \frac{1}{c} \underbrace{\left[v_jT^{ij}_{\text{Hall}}+P_{\not P} v^i-\frac{\partial P}{\partial \Omega_{nr}}\Omega_{nr}v^i \right]}_{j^i_{\epsilon \not P}}+ \mathcal{O}(1/c^3) \, , \\
T^{ij}_{\not P} &= \underbrace{-\tilde\eta_{nr} (\epsilon^{ik}\delta^{jl}+i\leftrightarrow j) V_{kl}}_{T^{ij}_{\text{Hall}}}+P_{\not P}\delta^{ij}+ \mathcal{O}(1/c^2)\, ,
\end{split}
\end{equation}
where $P_{\not P}=-\frac{\partial P}{\partial \mathcal{B}} \mathcal{B}-\frac{\partial P}{\partial \Omega_{nr}}\Omega_{nr}$ and $V^{kl}=\frac{1}{2}\left(\partial^k v^l+\partial^l v^k \right)$. Similar to the relation between the relativistic and non-relativistic dissipative viscosity coefficients, here we defined the non-relativistic Hall viscosity coefficient $\tilde\eta_{nr}=\tilde\eta/c$. By combining this with Eq. \eqref{NRlim}, the parity-violating modifications to the constitutive relations can be identified in the non-relativistic hydrodynamics. First, we find the non-relativistic Hall contribution to the stress tensor $T^{ij}_{\text{Hall}}$ known as the Hall viscosity \cite{Avron:1995fg, 1997physics..12050A, 2009PhRvB..79d5308R, 2011PhRvB..84h5316R} and the corresponding modification of the energy current of the form $j_{\epsilon, \text{Hall}}^i=v_j T^{ij}_{\text{Hall}}$. The non-relativistic Hall viscosity is dissipationless, since its contribution to the entropy production equation is  $\sim T^{ij}_{\text{Hall}} V_{ij}=0$. Second, the parity-violating magnetic field $\mathcal{B}$ and vorticity $\Omega_{nr}$ produce a shift of the pressure $P_{\not P}$ which leads to the corresponding additional terms in the stress tensor and the energy current. Finally, the vorticity $\Omega_{nr}$ also generates a shift in the non-relativistic energy density and a new contribution in the energy current.

The non-relativistic limit of the parity-violating terms in the $U(1)$ symmetry current $J^{\mu}$ in Eq. \eqref{E:T1J1L} reads
\begin{equation} \label{NRlimpvc}
 \begin{split}
J^{0}_{\not P} &=\frac{1}{c^2}\Big(\frac{1}{m}  \left[ \frac{\partial \Pi}{\partial \Omega_{nr}} \right]\epsilon^{ij} v_i\partial_j \ln T -\frac{\partial P}{\partial \mathcal{B}}\Omega_{nr}\Big)+\mathcal{O}(1/c^4)\, , \\  
J^{i}_{\not P} &=\frac{1}{c}\frac{1}{m}  \left[ \frac{\partial \Pi}{\partial \Omega_{nr}} \right]  \epsilon^{ij}\partial_j \ln T-\frac{1}{c^3}\Big(\frac{1}{m}j^{i}_{\tilde\epsilon}+\frac{\partial P}{\partial \mathcal{B}}\Omega_{nr} v^i\Big)+\mathcal{O}(1/c^5) \, .
\end{split}
\end{equation}
Here we introduced 
\beq \label{jite}
j^{i}_{\tilde\epsilon}=\underbrace{-\tilde{\lambda}_{nr}\epsilon^{ij}\partial_j T}_{j^i_{\tilde\epsilon,{th}}}\underbrace{+\frac{\partial \Pi}{\partial \mathcal{B}}\left[\epsilon^{ij}\mathcal{E}_j-\mathcal{B}v^i\right]}_{j^{i}_{\tilde\epsilon,{el}}}\underbrace{- \frac{\partial \Pi}{\partial \Omega_{nr}}  \partial_t \ln T \epsilon^{ij}v_j}_{j^i_{\tilde\epsilon,{v}}}\, .
\eeq
We introduced the non-relativistic thermal Hall conductivity coefficient 
\beq \label{kappanr}
\tilde{\lambda}_{nr}=\tilde{\kappa}_{nr}+\frac{v^2}{2T}\frac{\partial \Pi}{\partial \Omega_{nr}}+\dots ,
\eeq
with $\tilde{\kappa}_{nr}=\lim_{c\to\infty}\frac{\tilde{\sigma}}{c} \frac{(\epsilon_0+P_0)^2}{n_{0}^2 T}$ and $\dots$ denoting the terms originating from the sub-leading $O(1/c^3)$ term of the non-relativistic expansion of $\tilde\chi_T$. For the sake of simplicity they are not displayed explicitly. Similar to the parity preserving case, $\tilde{\kappa}_{nr}$ is assumed to be finite.

 By comparing with Eq. \eqref{current}, we are now ready to read off the additional parity-violating terms in the non-relativistic constitutive relations: 
 
 We start with the parity-violating corrections to the non-relativistic energy current. First, we identify the thermal Hall contribution $j^i_{\tilde\epsilon,th}$. In the literature, the Hall energy flow sourced by the temperature gradient is known as the Leduc-Righi effect. In condensed matter physics it is predicted to appear in various topological states of matter such as quantum Hall systems, chiral superfluids and topological insulators\cite{PhysRevB.61.10267,PhysRevB.85.045104}, where the thermal Hall conductivity $\tilde{\kappa}_{nr}$ is quantized. Second, we obtain the (anomalous) electromagnetic Hall energy current $j^i_{\tilde\epsilon,el}$ that originates from the non-relativistic limit of the $\tilde\chi_{E}$ term in the relativistic constitutive relation. Third, we find the energy current contribution $j^i_{\tilde\epsilon,v}$ that is perpendicular to the velocity field and proportional to the time derivative of the temperature $T$. 
  
  We now consider the parity-violating corrections to the non-relativistic charge current \eqref{NRlimpvc}. While the anomalous Hall charge current is absent in non-relativistic hydrodynamics,\footnote{Note that at $T=0$ any Galilean-invariant system has a vanishing Hall conductivity in the absence of the magnetic background field \cite{PhysRevB.61.10267}. To our best knowledge there exists no general argument for $T\neq 0$. Our result suggests that the findings of \cite{PhysRevB.61.10267} might be valid more generally for $T\ne 0$.}
  we find the charge current Hall response to the temperature gradient.\footnote{Similar current response arises in a two-dimensional electron liquid in an external magnetic field \cite{CHR97}.} The presence of this term in the charge current is disturbing since it does not appear in the leading $O(1/c)$ order term of $T^{0i}_{\not P}$. For the internal consistency of the non-relativistic limit, $J^i$ and $T^{0i}$ must agree at the leading order of the non-relativistic expansion. Note, however, that for a susceptibility that depends only on temperature, but not on the chemical potential, i.e.  $\frac{\partial \Pi}{\partial \Omega_{nr}}=f(T)$, the parity-violating modification is actually a curl and thus the non-relativistic conservation law of charge
\beq \label{one}
\partial_t \frac \rho m+\partial_i \left(\frac \rho m v^i+ \frac 1 m \frac{\partial \Pi}{\partial \Omega_{nr}}   \epsilon^{ij}\partial_j \ln T\right)=0\, ,
\eeq
is consistent with the conservation law of mass
\beq \label{two}
\partial_t \rho+\partial_i \left(\rho v^i\right)=0 \, ,
\eeq
obtained from the non-relativistic limit of $\partial_{\mu}T^{\mu 0}=F^{\nu\mu}J_\mu$.  
Since we found no general argument why this susceptibility should not depend on the chemical potential, we will state this as an assumption for the rest of this paper. In particular, we assume
\beq
\frac{\partial \Pi}{\partial \Omega_{nr}}=P_1+\frac{w_{nr,0}}{\rho_0}\rho_1=f(T),
\eeq
where $f(T)$ is some unspecified function of $T$ and we introduced $P_1$ and $\rho_1$ by the Taylor expansion
\beq
\begin{split}
P(\mu_{nr},T,\Omega_{nr})&=P_0(\mu_{nr},T)+P_1(\mu_{nr},T)\Omega_{nr}+\mathcal{O}(\Omega_{nr}^2), \\
\rho(\mu_{nr},T,\Omega_{nr})&=\rho_0(\mu_{nr},T)+\rho_1(\mu_{nr},T)\Omega_{nr}+\mathcal{O}(\Omega_{nr}^2).
\end{split}
\eeq
Since $\rho_1=\partial P_1/\partial \mu_{nr}$, we get the differential equation
\beq
P_1+\frac{w_{nr,0}}{\rho_0}\frac{\partial P_1}{ \partial \mu_{nr}}=f \, ,
\eeq
which has the solution
\beq
P_1(\mu_{nr},T)=C_1\exp\Big[-\int^{\mu_{nr}}_{\mu_{nr,0}} \frac{\rho_0(\tilde \mu, T)}{w_{nr, 0}(\tilde \mu, T)} d\tilde \mu \Big]+f(T),
\eeq
where $C_1=P_1(\mu_{nr,0},T)-f(T)$. We thus come to the conclusion that the internal consistency of the non-relativistic limit implies that the function $P_1(\mu_{nr}, T)$ can be fixed in terms of the function $f(T)$ and thermodynamics at $\Omega_{nr}=0$.

In general, one can always redefine the spatial part of the conserved current by adding a curl, i.e.,
\beq
j^i \to j^i+\epsilon^{ij}\partial_j F,
\eeq
where $F$ is an arbitrary scalar function of $t$ and $\mathbf{x}$. Since this does not affect the conservation law, one may wonder if the current can be defined in the unique way. It turns out that the current can be fixed by studying its coupling to external electromagnetic field.
To this end we write down the hydrodynamic equations of the parity-violating non-relativistic theory in the presence of electromagnetic sources
\beq \label{consP}
\boxed{
\begin{split}
&\partial_t \rho+\partial_i (\rho v^i)=0, \\
&\partial_t (\rho v^i)+\partial_j \Pi^{ij}_{tot}=\mathcal{E}^i j^0+\mathcal{B}\epsilon^{ij}j_j , \\
&\partial_t j_{\epsilon,tot}^0+\partial_i j_{\epsilon,tot}^i=\mathcal{E}^i j_i, \\
\end{split}
}
\eeq
where we introduced the total stress tensor, energy density and energy current
\beq 
\begin{split} \label{PiJtot}
\Pi^{ij}_{tot}&=\Pi^{ij} \underbrace{-\tilde\eta_{nr} (\epsilon^{ik}\delta^{jl}+i\leftrightarrow j) V_{kl}}_{T^{ij}_{\text{Hall}}}+P_{\not P}\delta^{ij}, \\
j_{\epsilon,tot}^0&=\epsilon_{nr} + \frac{1}{2} \rho v^2- \left[ \frac{\partial \Pi}{\partial \Omega_{nr}} \right]\epsilon^{ij} v_i\partial_j \ln T , \\
j_{\epsilon,tot}^i&=j_{\epsilon}^i+j_{\tilde\epsilon}^i+T^{ij}_{\text{Hall}}v_j+P_{\not P} v^i ,
\end{split}
\eeq
and the charge density and current
\beq
\begin{split}
j^0&=\frac{\rho}{m}, \\
j^i&=\frac{1}{m}\left[\rho v^i+ \frac{\partial \Pi}{\partial \Omega_{nr}}   \epsilon^{ij}\partial_j \ln T \right].
\end{split}
\eeq

In order to see how parity-violating effects change non-relativistic hydrodynamics we have to compare the equations \eqref{conserv} and \eqref{consP}. The first equation, i.e. the continuity equation, remains unchanged after allowing parity-violation. On the right-hand-side of the second equation in \eqref{consP} the external electromagnetic field couples to the charge current which now contains the Hall term discussed earlier in this section. On the left-hand-side of that equation the stress tensor receives two parity-odd contributions: one due to Hall viscosity $\tilde \eta$, the other due to a shift $P_{\not P}$ of the equilibrium pressure which depends on the magnetic field $\mathcal{B}$ and the vorticity $\Omega_{nr}$ (see definition below Eq.~\eqref{NRlimpv}). The third equation in \eqref{consP} now contains the total energy density which is corrected by the temperature gradient as seen from \eqref{PiJtot}. The total energy current $j^i_{\epsilon,tot}$ receives three corrections: one due to the presence of Hall viscosity, another again due to the shift $P_{\not P}$ of the equilibrium pressure, and the third correction due to the parity odd current $j_{\tilde\epsilon}^i$ defined in \eqref{jite}.

Finally, it is worthwhile to compare our finding from this section for the non-relativistic charge current with the result of LCDR obtained in Appendix \ref{sec:dimensionalReduction}. The general structure of the non-relativistic current looks similar, but there are some important differences. First, both methods give rise to the convective contribution to the current. Note, however, that while the charge density obtained in this section is independent of vorticity, it is directly proportional to $\Omega_{nr}$ within LCDR.  Second, both the calculation in this section and LCDR lead to  the parity-violating Hall contribution in the non-relativistic charge current. They are, however, not equivalent, since in the former case the Hall current flows perpendicular to the gradient of temperature, while in the latter case there is a Hall response to the gradient of the entropy per particle.

\section{Hydrodynamic Frames} \label{sec:hydroFrames}
Here, following closely the lucid presentation in \cite{Kovtun:2012rj}, we first discuss how to transform between hydrodynamic frames in relativistic hydrodynamics. Then we take the limit $c\to\infty$ and consider frame transformations in the non-relativistic regime.

In relativistic hydrodynamics the energy-momentum tensor and the current can be generally decomposed with respect to the vector $u^{\mu}$ as
\begin{eqnarray} \label{dec}
  &&   T^{\mu\nu} = \mathcal{E}  u^\mu u^\nu
  			    +\mathcal{P} \Delta^{\mu\nu}
			    +(q^{\mu}u^{\nu}+q^{\nu}u^{\mu})
			    +t^{\mu\nu} \,,  \\
  &&  J^\mu = \mathcal{N}  u^\mu +j^{\mu} \, ,
\end{eqnarray}
with the scalars $\mathcal{E}$, $\mathcal{P}$ and $\mathcal{N}$, transverse vectors $q^{\mu}$ and $j^{\mu}$ and the transverse symmetric and traceless tensor $t^{\mu\nu}$.

Since out of equilibrium neither of the hydrodynamic variable has a fundamental microscopic definition, one can perform their redefinitions known as frame transformations 
\begin{equation} \label{frama}
T\to T + \delta T, \quad \mu\to \mu + \delta \mu, \quad u^\mu\to u^\mu + \delta u^\mu .
\end{equation}
Note that the variations $\delta T$, $\delta \mu$ and $\delta u^{\mu}$ depend only on gradients of the hydrodynamic variables and thus vanish in equilibrium.\footnote{This becomes more subtle for parity-violating equilibria with finite vorticity. We do not consider this case in the present paper.} By definition, the frame transformation should not change the energy-momentum tensor $T^{\mu\nu}$ and the current $J^{\mu}$. For $\delta T$, $\delta \mu$ and $\delta u^{\mu}$ that are first order in gradients this implies
\beq \label{dc}
\delta q^{\mu}=-(\mathcal{E}+\mathcal{P})\delta u^{\mu}, \qquad \delta j^{\mu}=-\mathcal{N}\delta u^{\mu},
\eeq
while the rest of the coefficients in the decomposition \eqref{dec} remain invariant to first order. It is clear from Eq. \eqref{dc} that 
\beq
l^{\mu}=j^{\mu}-\frac{\mathcal{N}}{\mathcal{E}+\mathcal{P}} q^{\mu} \, ,
\eeq
is invariant with respect to the frame redefinition.

In practice, one can take advantage of the frame ambiguity  and pick up a particular frame that is suitable for a given problem. In particular, in parity-invariant hydrodynamics it is convenient to choose $T$ and $\mu$ such that $\mathcal{E}=\epsilon$ and $\mathcal{N}=n$, where $\epsilon$ and $n$ are energy density and charge density in equilibrium. As for the choice of $u^{\mu}$, there are two popular choices of frames: In the Eckart frame one has $j^{\mu}=0$, i.e.,  there is no charge current in the local rest frame. Alternatively, in the Landau frame $q^{\mu}=0$ and thus no heat is transported in the local rest frame.

By taking the non-relativistic limit of Eq. \eqref{frama} we obtain that for $v/c\ll 1$ the frame transformations are parametrized by\footnote{Although the spacetime velocity $u^{\mu}$ has one more component than the spatial velocity $v^i$, the number of independent frame transformations is actually the same in the relativistic and non-relativistic regimes. Indeed, due to the normalization condition of the relativistic velocity, $\delta u^{\mu}$ must be transverse, i.e. $\delta u^{\mu} u_{\mu}=0$, and thus $\delta u^{\mu}$ has only two independent components in two spatial dimensions.}
\begin{equation} \label{framaNR}
T\to T + \delta T, \quad \mu_{nr}\to \mu_{nr} + \delta \mu_{nr}, \quad v^i\to v^i + \delta v^i .
\end{equation}
This is natural, since $T$, $\mu_{nr}$ and $v^i$ are the hydrodynamic variables of non-relativistic hydrodynamics.  As before, the variations in Eq. \eqref{framaNR} should vanish in equilibrium.

As the first example, we consider the transformation of parity-invariant hydrodynamics from the Landau to the Eckart frame in the non-relativistic regime. It is worth emphasizing first that in the parity-preserving hydrodynamics the difference between the Landau and Eckart frame is tiny in the non-relativistic regime because $\delta v^{i}=O(1/c^2)$. Indeed, since at low velocities most of the total energy is stored in the rest energy of the particles, the energy and the charge currents are almost collinear for a one-component fluid.\footnote{This means that the transformation from the Landau to Eckart frame does not belong to the set of transformations which could be realized in the non-relativistic hydrodynamic theory.}
This particular frame transformation is achieved by the substitution
\beq \label{LtE}
v^i\to v^i+\frac{1}{c^2 \rho}j^{i}_{\epsilon, th}+\mathcal{O}(1/c^4) \, ,
\eeq
which induces
\beq \label{LtE1}
\rho \to \rho+\mathcal{O}(1/c^4) \, ,
\eeq
since we defined $\rho=\Gamma n m$, 
and the $\Gamma$ factor is defined by the square of the velocity, $v^2$. All other identifications we have made previously in order to relate relativistic quantities to non-relativistic ones are only affected at order $1/c^4$ or higher, as can be checked by explicit computation.
By substituting transformations \eqref{LtE}, \eqref{LtE1} into Eq. \eqref{current} we  find that the Eckart conditions
\beq \label{Eckart}
\begin{split}
J^0&=\frac{\rho}{m} +\mathcal{O}(1/c^4), \\
J^i&=\frac{1}{c}\frac{\rho v^i}{m} +\mathcal{O}(1/c^5) \, ,
\end{split}
\eeq
are satisfied (up to higher orders in $1/c$ expansion), i.e., the current is purely convective in the non-relativistic regime. At the same time, the form of the energy-momentum tensor is modified by
\begin{equation} \label{NRlimdelta}
 \begin{split}
\Delta T^{00} &=\mathcal{O}(1/c^2)\, , \\
\Delta T^{0i} &= \frac{1}{c}j^{i}_{\epsilon, th}+\mathcal{O}(1/c^3)\, , \\
\Delta T^{ij} &= \mathcal{O}(1/c^2)\, ,
\end{split}
 \end{equation}
in other words in the Eckart frame the thermal conductivity term $j^{i}_{\epsilon, th}$ is a part of the energy current. Since the frame transformation does not modify Eq. \eqref{Son}, we find the textbook non-relativistic dissipative hydrodynamics also as the limit of the relativistic hydrodynamics formulated in the Eckart frame.

Let us now look at the transformation from the magnetovortical to the Eckart-like frame in the non-relativistic regime of the parity-violating hydrodynamics. This is done by the frame transformation
\beq \label{MVtE}
\begin{split}
v^i&\to v^i-\frac{1}{\rho}\frac{\partial \Pi}{\partial \Omega_{nr}}\epsilon^{ij}\partial_j \ln T+\frac{1}{c^2}\frac{1}{\rho}[j^{i}_{\epsilon, th}+ j^{i}_{\tilde \epsilon}]+\mathcal{O}(1/c^4), \\
T&\to T+\frac {1} {c^2} \frac{\frac{\partial P}{\partial \Omega_{nr}}\Omega_{nr}}{ \frac{\partial \rho}{\partial T}}+\mathcal{O}(1/c^4),
\end{split}
\eeq
which together (due to the identification $\rho=\Gamma n m$ and the relation $\delta \rho=(\partial \rho/ \partial T) \delta T $) induce
\beq
\begin{split}
\rho&\to \rho-\frac{1}{c^2}\Big( \left[ \frac{\partial \Pi}{\partial \Omega_{nr}} \right]\epsilon^{ij} v_i\partial_j \ln T -\frac{\partial P}{\partial \Omega_{nr}}\Omega_{nr}\Big)+\mathcal{O}(1/c^4) .
\end{split}
\eeq
By substituting these transformations into Eqs. \eqref{current}, \eqref{NRlimpvc} we find that the Eckart conditions \eqref{Eckart} 
are satisfied.
We observe that if parity is broken, the variation of the velocity field is not small anymore, but contains the term that is $ O(c^0)$. In addition, in contrast to the parity-invariant sector the frame transformation includes the transformation of the temperature $T$. 

We find that in the Eckart-like frame the terms $j^i_{\epsilon, th}$ and $j_{\tilde \epsilon}^i$  from Eq.  \eqref{jite} contribute to the energy current. In addition, if the susceptibility $\frac{\partial \Pi}{\partial \Omega_{nr}}$ is non-vanishing, we find corrections to the energy density, momentum current, energy current and stress tensor. We refrain from writing out these lengthy expressions here.

As a general lesson our results suggest that we are always free to start in any relativistic hydrodynamic frame, take the non-relativistic limit, and we will arrive at the corresponding non-relativistic frame. However, there are frames between which there exists no simple transformation in the non-relativistic theory itself. 
One example for this is the transformation from Landau to Eckart frame presented previously. 
The difference between these frames becomes visible if one includes relativistic corrections to the non-relativistic hydrodynamics. Hence it seems necessary to start in the relativistic parent theory, make the frame transformation there, and then perform the $1/c$-expansion.
In this process one has to include the transformations which are hidden in thermodynamic quantities such as the particle density $\rho$, as seen from the examples above.

\section{Conclusion}\label{sec:conclusions}
In this paper the non-relativistic hydrodynamics of a parity-violating two-dimensional normal fluid was constructed. Our main results are the non-relativistic constitutive equations \eqref{NRlimpv}, \eqref{NRlimpvc}, and the non-relativistic conservation equations \eqref{consP} including various parity-violating contributions. We have arrived at these results making two assumptions: (i) We assumed that thermodynamic quantities only depend on the general coordinate invariant combination $\mathcal{B}+m \Omega_{nr}$. The physical motivation for this assumption is the  fact that our fluid consists of single species of particles which carry both charge and mass inseparably. If simultaneously subjected to vorticity and magnetic field, these particles will experience a Lorentz force whenever they experience a Coriolis force. (ii) We have assumed that the susceptibility $\partial\Pi/\partial \Omega_{nr}$ can only depend on the temperature $T$, but not on the chemical potential $\mu$. One might speculate that our assumption (i) was too restrictive and that assumption (i) forced us to make assumption (ii). But at this time we have no indication in favor of or against this speculation. Therefore we have to conclude that our results apply to a subset of all possible non-relativistic parity-violating fluids, namely those which satisfy assumptions (i) and (ii).

As advertised in the introduction, our derivation implies that a particular contribution to the energy current perpendicular to a temperature gradient can be expressed in terms of thermodynamic quantities as seen in equation \eqref{PiJtot}. Provided the energy density, pressure, and particle density are known, our approach predicts the linear response of the system to a temperature gradient. Similarly, given fluctuations in the magnetic field and in the vorticity, the resulting shift in the pressure is predicted to be given by $P_{\not P}=-\frac{\partial P}{\partial \mathcal{B}} \mathcal{B}-\frac{\partial P}{\partial \Omega_{nr}}\Omega_{nr}$.

One open question remaining is if we indeed get a more restricted version of non-relativistic hydrodynamics than one would obtain from merely requiring Galilean invariance alone. In order to find the answer, we would need to construct the constitutive relations for parity-violating hydrodynamics invariant only under Galilean symmetry. We know of no systematic construction principle for that case, but finding it would be a worthwhile task for future work.

Another possible application of our method and extension of our work should be the systematic computation of relativistic corrections to non-relativistic hydrodynamics. In the spirit of corrections to an effective field theory, this would require to retain terms of higher order in our expansion in the inverse speed of light. In this work we have not carried out this analysis explicitly, but we deem it worthwhile for future work.

Our non-relativistic parity-violating normal fluid hydrodynamics should give an effective description of the physical systems mentioned in the introduction. However, arguably the most interesting parity-violating non-relativistic quantum fluids can not be described by the present theory. Among the most interesting we mention:
\begin{itemize}
\item {\bf Chiral $p_x+ip_y$ superfluids:}  In chiral superfluids spontaneous symmetry breaking of continuous symmetries leads to parity breaking. The superfluid hydrodynamics at vanishing temperature is governed by the dynamics of Goldstone boson(s). These degrees of freedom are absent in our present description.
\item {\bf Non-relativistic quantum Hall fluids:} Finite background magnetic field ${\mathcal B}$ is a necessary ingredient of the quantum Hall physics. In contrast, in this work we always expanded around the equilibrium with vanishing background magnetic field. 
\end{itemize} 
For recent developments of the parity-violating hydrodynamics of chiral $p_x+i p_y$ superfluids and quantum Hall systems we refer to \cite{Hoyos2013, Abanov2012,Wiegmann2012, Son2013,Golkar:2013gqa}. 
It would be very interesting to develop the hydrodynamic description of the above-mentioned parity-violating quantum fluids starting from the relativistic version of hydrodynamics in the future.

Furthermore, it would be worthwhile testing our assumptions and results explicitly within a holographic model making use of the gauge/gravity correspondence~\cite{Maldacena:1997re} developed for non-relativistic field theories in~\cite{Janiszewski:2012nf,Janiszewski:2012nb}. One concrete example for a non-relativistic holographic model derived from string theory is the system discussed in~\cite{Ammon:2012je}.

\section*{Acknowledgements}
We thank S. Janiszewski, K. Jensen, A. Karch, P. Kovtun, M. Rangamani, A. Ritz, D. T. Son, M. Stephanov, and R. Thomale  for both helpful and inspiring discussions. MK is currently supported by the US Department of Energy under contract number DE-FGO2-96ER40956. The work of S.M.~was supported by US DOE Grant No.~DE-FG02-97ER-41014.

\begin{appendix}
\section{Light-Cone Dimensional Reduction} \label{sec:dimensionalReduction}
Here we refrain from the systematic study and as a relativistic parent theory take the ideal hydrodynamics in 3+1 dimensions supplemented with the anomalous current
\beq
J^{\mu}_{an}=\frac{\xi}{2}\epsilon^{\mu\nu\rho\sigma}u_{\nu}\partial_{\rho}u_{\sigma}.
\eeq
As shown in \cite{Son:2009tf}, this current is required by triangle anomalies and the second law of thermodynamics. It is responsible for the chiral vortical effect, i.e. chiral separation in a relativistic rotating fluid. The anomalous current is present even at the vanishing density $n=0$, when there is no convection. In that case $\xi=C_{\mu}\mu^2+C_T T^2$ with $C_{\mu}$ and $C_T$ determined by the anomalies.

LCDR of the anomalous current $J^{\mu}_{an}$ generates the non-relativistic charge density $q_{an}$ and charge current $j^{i}_{an}$. In particular, we find\footnote{In the light-cone coordinates for the metric we use the convention $\eta_{+-}=\eta_{-+}=-1$ and $\eta_{ij}=\delta_{ij}$. In addition $\epsilon^{+-ij}=\epsilon^{ij}=\epsilon_{ij}$ with $\epsilon_{12}=1$.}
\beq
\begin{split}
q_{an}&\equiv j^{+}=-\xi (u^+)^2 \Omega_{nr}= -\frac{\xi}{4}\frac{\rho}{w_{nr}} \Omega_{nr}, \\
j^i_{an}&=\underbrace{q_{an} v^i}_{\text{convection}}-\underbrace{\frac{\xi}{2}\epsilon^{ij}\Big[\frac{(u^+)^2}{\rho}\partial_j P+\frac{1}{u^+} \partial_j u^{+} \Big]}_{\text{Hall}}
=q_{an} v^i-\frac{\xi}{4}\epsilon^{ij}\underbrace{\Big[\frac{\partial_j \rho}{\rho}-\frac{\partial_j \epsilon_{nr}}{w_{nr}} \Big]}_{H_j},
\end{split}
\eeq
where $\Omega_{nr}\equiv\epsilon^{jk}\partial_j v_k$ and $u^+=\sqrt{\frac{\rho}{2w_{nr}}}$. Here following the main part of our paper $\rho$, $\epsilon_{nr}$, $w_{nr}$ and $P$ represent the non-relativistic mass density, internal energy density, enthapy density and pressure, respectively. In the derivation we used identifications between ideal relativistic and non-relativistic hydrodynamics variables found in \cite{Rangamani:2008gi}, the momentum conservation equation of the ideal non-relativistic hydrodynamics and the kinematic identity $-\Omega_{nr} v^i=\epsilon^{ij}v^k (\partial_j v_k-\partial_k v_j)$. 

Using thermodynamic arguments we can simplify our result for $H_j$. Indeed, in the canonical ensemble the first law of thermodynamics
\beq
d E_{nr}=T dS-p dV \, ,
\eeq
can be rewritten in the intensive form as
\beq
\frac{T dS}{V}=d\epsilon_{nr}-\frac{w_{nr} d \rho}{\rho}.
\eeq
This allows us to write
\beq
H_j =-\frac{T}{V w_{nr}}\partial_j S=-\frac{\rho T}{m w_{nr}}\partial_j S_p.
\eeq

We observe that the non-relativistic charge density is a kind of angular momentum density since it is generated by a rigid rotation with a constant $\Omega_{nr}$. The charge current $j_{an}^i$ splits naturally into the convective and the Hall parts. The Hall contribution is proportional and perpendicular to the gradient of the entropy per particle $S_p=\frac{S}{N}$.

For $n=0$, due to the anomaly the current $J^{\mu}_{an}$ is conserved only if $E^{\mu}B_{\mu}=0$, where $E^{\mu}=F^{\mu\nu}u_{\nu}$ and $B^{\mu}=\frac{1}{2}\epsilon^{\mu\nu\rho\sigma}u_{\nu}F_{\rho\sigma}$. In this particular case LCDR leads to the conservation of the non-relativistic charge, i.e.
\beq
\partial_{\mu}J^{\mu}_{an}=0 \rightarrow \partial_+ q_{an}+\partial_i j^i_{an}=0.
\eeq

For a systematic study of the parity-violating hydrodynamics from LCDR we refer to \cite{Brattan:2010bw}.

\end{appendix}

\bibliography{hydrobib_2plus1}{}
\end{document}